# A Trust Based Congestion Aware Hybrid Ant Colony Optimization Algorithm for Energy Efficient Routing in Wireless Sensor Networks (TC-ACO)

Arpita Chakraborty[1], Srinjoy Ganguly[2], Mrinal Kanti Naskar[3], Anupam Karmakar[4]

[1] Dept. of Electronics and Communication Engineering, Techno India, Salt Lake, Kolkata 700091, India
[2, 3] Dept. of Electronics and Telecommunication Engineering, ADES Lab, Jadavpur University, Kolkata 700032, India
[4] University of Calcutta, Dept. of Electronic Science, Kolkata 700009, India
[1]carpi.technoindia@yahoo.com, [2]srinjoy_ganguly92@hotmail.com, [3]mrinalnaskar@yahoo.co.in, [4]anupamkarmakar@yahoo.co.in

*Abstract* - Congestion is a problem of paramount importance in resource constrained Wireless Sensor Networks, especially for large networks, where the traffic loads exceed the available capacity of the resources. Sensor nodes are prone to failure and the misbehavior of these faulty nodes creates further congestion. The resulting effect is a degradation in network performance, additional computation and increased energy consumption, which in turn decreases network lifetime. Hence, the data packet routing algorithm should consider congestion as one of the parameters, in addition to the role of the faulty nodes and not merely energy efficient protocols. Unfortunately most of the researchers have tried to make the routing schemes energy efficient without considering congestion factor and the effect of the faulty nodes. In this paper we have proposed a congestion aware, energy efficient, routing approach that utilizes Ant Colony Optimization algorithm, in which faulty nodes are isolated by means of the concept of trust. The merits of the proposed scheme are verified through simulations where they are compared with other protocols.

## I. Introduction

Wireless Sensor Networks (WSNs) have immense potential for a variety of applications in diverse fields, where employment of human beings is not feasible. It consists of a large number of randomly deployed sensor nodes, which are battery-powered and resource constrained in terms of limited energy, circumscribed computational and communication capability, bounded memory and processing speed [1 - 2]. Since the turn of the 21st century, researchers throughout the globe have proposed various energy efficient routing algorithms to maximize the network lifetime of WSNs. Yet, a scope for improvement does exist. Sensor networks are generally operated in an idle mode and then suddenly become active in response to the detected event [3]. When a large number of sensor nodes become active, so as to transport data to the base station (BS), packet collisions and network level congestion are quite probable. Buffer overflow occurs due to limited buffer size of the nodes causing packet drops which in turn decreases network performance and throughput. Moreover, inexpensive sensor nodes are prone to failure and the faulty nodes create various security threats which aggravate the problem of congestion by diffusing useless packets, flooding with fake messages, intermittent jamming and/or retransmitting the same message several times. The resulting effect is redundant computation and communication, causing wastage of energy resources and a decrease in network lifetime. If the problematic faulty nodes are eliminated from the data routing path, the congestion of the network can be minimized which in turn enhances energy efficiency and the lifetime of the network. To address this challenge, we propose a congestion aware routing scheme based on the Ant Colony Optimization (ACO) algorithm in which faulty nodes, known as the malicious nodes, are detected and isolated by utilizing a trust-based framework [4]. Trust management is a relatively new idea which is used these days for detecting faulty nodes in order to establish a trustworthy data routing path from the source node to the BS. The concept of trust is basically borrowed from the human society, in which the sensor nodes monitor the behavior of their neighbors during previous data transfer operations through these nodes, on the basis of some parameters known as the Trust Metrics (TM) [5]. The congestion of the trusted node is computed by estimating the free space in the buffer queue of the node. In the proposed TC-ACO algorithm, ACO is utilized which factors in distance along with trust and congestion for energy efficient, trustworthy, optimal routing in WSNs.

The ACO, initially proposed by Marco Dorigo [6-8], is based on the behavior of real ants, while they search for their food in short routes from their nest to the location of the food. While travelling, an ant deposit pheromone in its path and the intensity of the pheromone decreases over time due to evaporation. The ants following shorter paths are expected to return earlier through the same path, compared to the ants in longer paths. Hence, the amount of pheromone deposited in the shortest path is more than that of the other paths. The new ants are subjected to follow the shortest path having more pheromone. In this way, the pheromone deposition in the shortest path increases whereas the other paths are lost due to lack of pheromone. Finally, all the ants follow the shortest path. In the proposed scheme data packets, considered as artificial ants, are launched from the source node and find the optimal route towards the destination node in each cycle.

The rest of the paper is organized as follows: In section II, some existing related protocols are discussed. The proposed algorithm is presented in section III, the simulation results and comparisons with other existing protocols are discussed in section IV and finally section V concludes the paper.

## II. RELATED WORKS

Trust based congestion aware routing in WSNs is a relatively new research topic and has not been addressed in literature to a great extent. Although a lot of energy efficient routing protocols are available, most of them do not consider network security, role of the faulty nodes and the problem of congestion in their ambit. For example, a multi path routing protocol based on dynamic clustering and ACO, is described in MRP [9], which improves the efficiency of data aggregation, thereby reducing the energy consumption. The routing protocols with trust management are described in TRANS [10] and TILSRP [5]. However all the afore - mentioned protocols do not address the problem of congestion. Congestion and trust both are discussed in [11 - 13]. In the FCC protocol [11], Zarei et al. propose a Fuzzy based trust estimation for congestion control in WSNs. FCCTF protocol [12] is basically a modification of FCC, in which the Threshold Trust Value is used for decision making. Our previous work is the TFCC protocol [13], in which traffic flow from the source to sink is optimized by implementing the Link State Routing Protocol which provides improvement in network throughput.

## III. PROPOSED WORK

In this paper, we have presented a novel trust based congestion aware energy efficient routing scheme for WSNs in which the ACO algorithm is utilized to maximize the network lifetime. We consider random deployment of sensor nodes in the sensor field under free space propagation. The proposed algorithm works in two stages. In stage 1, the trust values and the congestion statuses of the nodes are calculated and thereby, the trust-congestion metric is formed. In stage 2, the ACO algorithm, which utilizes the trust-congestion metric and the distance metric, is implemented for data packet routing from source node to base station. The detailed operation of each stage is described below.

### A. Stage 1

In the proposed algorithm, stage 1 detects the mishaviour of the sensor nodes using the concept of trust. The trusted nodes (having trust value above some pre defined threshold level) are identified and congestion status are computed accordingly. The trust congestion metric is generated for the trusted nodes, also called valid nodes, for the data packet routing algorithm which is implemented in the next stage. The malicious nodes having trust value below the threshold level are not considered for data packet routing, due to which the congestion metric isn't computed for such nodes. This causes a reduction in the computation overhead and thereby enhances battery life time.

*1) Trust Computation :*

The trust value of node i upon node j is calculated on the basis of three commonly used trust metrics namely, remaining node energy ($N_e'$), packet transmission ratio ($P_{TR}'$) and packet latency ratio ($P_L'$). All the parameters are normalized so that the values belong to the range: [0,1]. $P_{TR}'$ is defined as the ratio of the number of acknowledgement received from node j to the total number of packets sent from node i to node j. $P_L'$ is the ratio of the latency of node j to the mean latency of the other nodes except node j, when data packets are transmitted from node i. $N_e'$ is defined as the average energy of the node i and node j. if $E_i$ and $E_j$ are the existing energy value of node i and node j respectively, then $N_e' = (E_i + E_j) / 2$. The energy of a sensor node should be greater than or equal to the threshold value of $E_{th}$ for transmitting data packets to its one-hop neighbor in the radio communication range.

Mathematically, the net trust of node i upon node j is calculated by the formula represented as :

$$T_{ij} = \frac{A_1 * N_e' + A_2 * P_{TR}' + A_3 * P_L'}{A_1 + A_2 + A_3} \quad (1)$$

where $A_1$, $A_2$ and $A_3$ are the corresponding weights used for $N_e'$, $P_{TR}'$ and $P_L'$ respectively such that $A_1, A_2, A_3 \in [0,1]$. A predefined trust threshold value ($T_{TH}$) is set on the basis of the application of the sensor networks [11]. If $T_{ij} > T_{TH}$, the link between the node i and j is called a trustworthy link. Similarly, if $T_{ij} < T_{TH}$, the link is termed as an untrusted link. The nodes having no trustworthy link are called malicious nodes and those with at least one trusted link are called trusted nodes (valid nodes) that can take part in data packet routing.

*2) Estimation of Node Congestion :*

The congestion level of a valid node is estimated with the help of the parameter known as the Congestion Index. It is assumed that each node maintains a queue for storing data packets in its buffer. As packets are transmitted from a particular node serially towards the next node, buffer space is cleared and the packets waiting in the queue go to the empty buffer space of the node. When the packet received rate of the node is greater than the packet transmission rate, queue length increases, buffer overflows, congestion level of the node increases. If a node is not able to clear the data packet in its queue, then it waits for a certain number of pre-defined cycles (say, $WC_{max}$) and holds the packets in each cycle until the packets are finally dropped (at the end of $WC_{max}$ cycles).

The Congestion Index of the $k^{th}$ node is computed by the equation given as:

$$CI_k = \frac{\overline{r}_{in}^k + Q^k(c-1) - \overline{r}_{out}^k}{\overline{r}_{in}^k + Q^k(c-1)}, \quad (2)$$

where $Q^k(c-1)$ is the empty space left in the queue of the kth node till $(c-1)^{th}$ cycle. The parameters $\overline{r}_{in}^k$ and $\overline{r}_{out}^k$ are defined as:

$$\overline{r_{in}^k} = \frac{\sum_{i=1}^{c-1}(N_{i,k}^A)}{c-1} \quad (3)$$

$$\overline{r_{out}^k} = \frac{\sum_{i=1}^{c-1}(N_{i,k}^B)}{c-1} \quad (4)$$

$N_{i,k}^A$ = Number of packets forwarded to the $k^{th}$ node in the $i^{th}$ cycle.

$N_{i,k}^B$ = Number of packets forwarded by the $k^{th}$ node to the other nodes in the $i^{th}$ cycle.

The congestion index of each trusted node, which is calculated by using equation (2), represents the node level congestion of the WSN. It is calculated dynamically at regular intervals, depending upon the application of the network.

*3) Computation of Trust Congestion Metric ($TC_{ij}$):*

The Trust Congestion Metric (TCM) of each trusted nodes, also called as the valid node, is computed by the equation:

$$TC_{ij} = \alpha * CI_j + (1-\alpha) * T_{ij} \quad (5)$$

where node *i* and node *j* are considered as the source node and the destination node, respectively. $CI_j$ is the Congestion Index of the destination node and $T_{ij}$ is the trust value of source node i upon the destination node j. The constant α is denoted as Trust Congestion Coefficient which belongs to [0,1].

*B. Stage 2*

The proposed TC –ACO algorithm stage 2 implements the data routing protocol using Ant Colony Optimization [6] – [8].

The probability $P_{ij}$ for transmission of data packets in optimal route from node i in level L to node j in level (L+1) is given by equation (6).

$$P_{ij} = \frac{\left(TC_{ij}\right)^{\beta_1} \cdot \left(\frac{1}{d_{ij}}\right)^{\beta_2} \left(\tau_{ij}\right)^{\beta_3}}{\sum_{\forall k}\left(TC_{ik}\right)^{\beta_1}\left(\frac{1}{d_{ik}}\right)^{\beta_2}\left(\tau_{ik}\right)^{\beta_3}} \quad (6)$$

where ' k' represents all the valid nodes in level (L+1).

We consider random deployment of sensor nodes in the entire sensor field at the various levels as depicted in Fig. 1, denoted by level 0, 1, 2 …level L, (L+1)…. level (r-1), level r respectively. The source node is considered as a level 0 node. All nodes within the one hop neighbor of the source node in the radio communication range are denoted as the level 1 nodes. Similarly, all nodes within the one hop neighbor of the level 1 in the radio communication range are called the level 2 nodes and so on.

At the end of the cycle c, $\tau_{ij}$ is updated as:

$$\tau_{ij}(c) = (1-\rho) \cdot \tau_{ij}(c-1) + \frac{N_{ij}}{d_{ij}} \quad (7)$$

The list of the variables used in afore - mentioned equations, are described in Table I.

TABLE I
LIST OF VARIABLES

| Variable Name | Description |
|---|---|
| $TC_{ij}$ | Trust Congestion Metric between node i and j |
| $\tau_{ij}$ | Pheromone concentration on the link connecting nodes i and j |
| $d_{ij}$ | Distance between nodes i and j |
| ρ | Evaporation constant ( ρ ∈ [0,1] ) |
| $N_{ij}$ | Nmber of data packets transmitted between nodes i and j in cycle (c-1) |
| $\beta_1, \beta_2$ and $\beta_3$ | Constant parameters with each ∈ [ 0,1] |

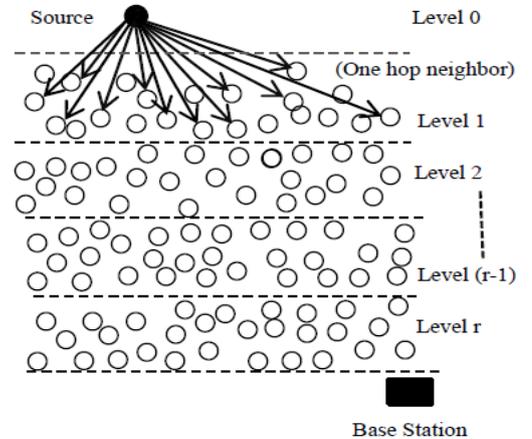

Fig. 1. Random Deployment of Sensor Nodes at the Various Levels

The data packet routing algorithm used in the proposed TC-ACO scheme is represented in Fig. 2.

```
Input: Trust Threshold Level, Trust Congestion
Metric
Output: Optimal Route

Begin
For each node i in Lth level,
  do
 Step 1: Find all valid node "j" in level
(L+1). The nodes satisfying the condition
T_ij > T_th are valid nodes
 Step 2: Compute probability of packet
transmission from node "i" to node "j" defined
as P_ij
 Step 3: Arrange the valid nodes in descending
order based on the value of P_ij and store in
matrix X.
 Step 4: Initialize m =1
    While (E_i ≥ E_th && m ≤ size(X))
     // Forward packet from node i to X(m)
     // Update energy of both nodes.
 If (Queue (X(m)^th node) is full OR (E_X(m) < E_th ))
    m = m+1;
 end –if
    end – while
   end – do
End
```

Fig. 2. Data Packet Routing Algorithm

## IV. SIMULATION RESULTS

In this section, the merits of the proposed TC-ACO scheme have been investigated through MATLAB simulations. We have considered an arbitrary network, comprising of 50 homogeneous nodes deployed randomly into a field of dimension 200 m * 200 m. The distances of the nodes from the base station (BS) are taken constant throughout the experiment. It is also assumed that the nodes are connected within the network in different levels as per the first order radio model [14]. Fig. 1 shows the graphical view for the packet routing, when 20 data packets are transmitted from the source node, marked as 27, towards the BS. The packets are routed in five different paths, of which 12 packets are transmitted through the optimal route, as obtained by ACO algorithm, which is marked in red color in Fig.3. During the next cycle of data transfer, the optimal route would be selected dynamically on the basis of the trust congestion metric of the corresponding nodes.

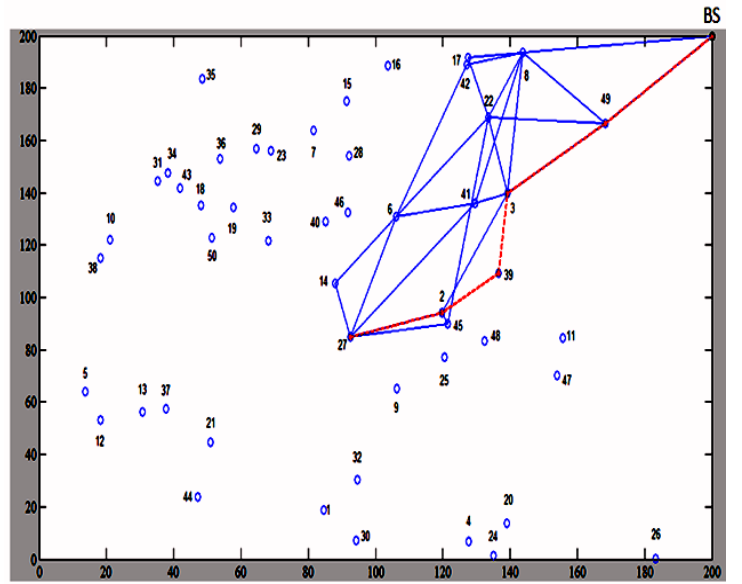

Fig. 3. Routing of packets (20 in number) from source node 27 to the BS

The performance of the proposed TC-ACO algorithm has been tested through rigorous MATLAB simulations, by varying node energy and trust threshold level. In this paper, we have presented the results obtained with initial node energy of 1.0 Joule per node and trust threshold level equal to 0.5. The proposed TC-ACO protocol is compared with the existing algorithms such as TRANS [10], MRP [9] and TFCC [13] respectively. Table II represents the number of rounds verses percentage of dead nodes with the initial energy of 1.0 Joule/node for the above mentioned protocols.The comparison of the proposed TC-ACO scheme with the other similar protocols has been shown graphically in Fig. 4 and Fig. 5 respectively. The percentage of dead nodes are plotted along the x axis whereas the number of rounds are ploted along the y axis of the graphs shown in Fig. 4 and Fig. 5. The simulation and experimental results indicate that the TC-ACO scheme provides higher network lifetime compared to other similar protocols and thereby outperform its peers.

The experimental results of the proposed TC-ACO algorithm are quite justified since TRANS and MRP do not consider additional energy consumption due to congestion. In TFCC and proposed TC-ACO algorithm, trust and congestion both are considered but data routing in the previous one is based on the Link Stare Routing Protocol whereas that in the proposed scheme is done through ACO.

TABLE II

PERFORMANCE ANALYSIS AT INITIAL ENERGY 1.0 JOULE / NODE

| Number Of Rounds | Protocols | Percentage of Dead Nodes | | | | | | |
|---|---|---|---|---|---|---|---|---|
| | | 1% | 10% | 20% | 30% | 40% | 50% | 60% |
| | TRANS | 1965 | 2132 | 2342 | 2596 | 2701 | 2910 | 3197 |
| | MRP | 2087 | 2221 | 2378 | 2601 | 2895 | 3020 | 3304 |
| | TFCC | 2223 | 2365 | 2455 | 2673 | 2812 | 3108 | 3276 |
| | TC-ACO | 2406 | 2677 | 2818 | 2997 | 3108 | 3285 | 3566 |

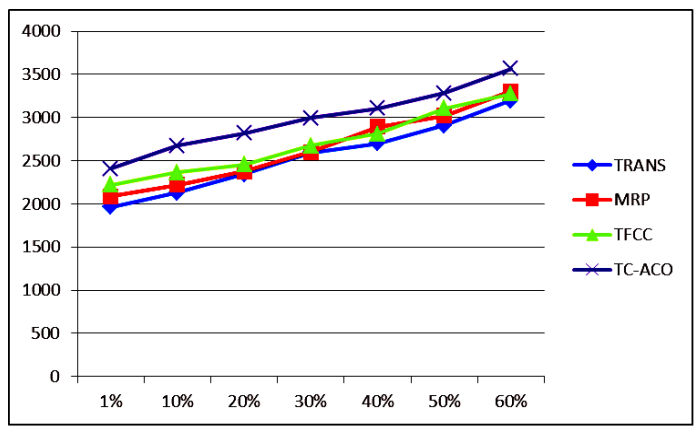

Fig. 4. Performance Analysis of the Protocols with Initial Energy 1.0 J/Node

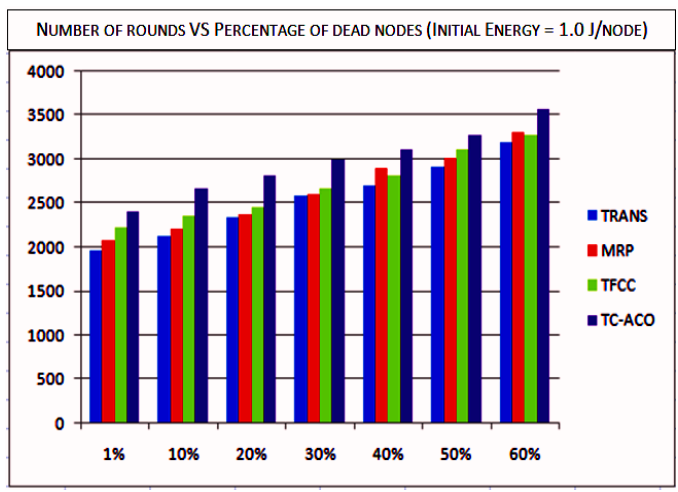

Fig. 5. Bargraph Analysis of the Protocols with Initial Energy 1.0 Joule/Node

## VI. CONCLUSION

In this paper we have presented a novel trust based congestion aware routing algorithm for WSNs in which the optimum route for data packet transfer is dynamically selected by the TC-ACO algorithm, on the basis of the trust, congestion level and inter-nodal distance of the sensor nodes. Our protocol considers the impact of the misbehavior of the faulty nodes on network congestion and thereby minimizes its effect during data packet routing. It is a very relevant research topic because in large WSNs, especially for Wireless Multimedia Sensor Networks, congestion is practically unavoidable. The simulation and experimental results indicate that the proposed TC-ACO algorithm provides the longest network lifetime compared to the other similar routing protocols. However, the proposed protocol is tested only on a small network. So, we need to test its adaptability to larger networks consisting of large number of nodes deployed over larger areas. In the future, we would also like to develop hardware with Iris motes and try to implement the same on TinyOS under various conditions.

REFERENCES

[1] I.F.Akyildiz, W.Su, Y.Sankarasubramaniam and E.Cayirci, " Wireless Sensor Networks: A Survey", Computer Networks, Elsevier, Vol. 38, March 2002, pp 393-422.
[2] D.Culler, D.Estrin, M.Srivastava, "Overview of Sensor Networks", IEEE Computer Society, August 2004.
[3] C.Y.Wan, S.B.Eisenman and A.T.Campbell, "CODA: Congestion Detection and Avoidance in Sensor Networks", SenSys' 03, Los Angeles, USA, pp. 266-279, ACM, Nov 2003.
[4] Mohammad Momani, Ph.D thesis on "Bayesian methods for modeling and management of Trust in Wireless Sensor Networks", University of Technology, Sydney, July, 2008.
[5] A. Raha, M.K. Naskar, S.S. Babu, Omar Alfandi, and D. Hogrefe, " Trust Integrated Link State Routing Protocol for Wireless Sensor Network (TILSRP), proceedings of 5th IEEE ANTS, Dec 2011.
[6] M. Dorigo, "Optimization, Learning and Natural Algorithms",PhD thesis, DEI, Politecnico di Milano, Italy (1992).
[7] M. Dorigo, G.Di Caro, "Ant Colony Optimization: a new meta-heuristic proceedings of the 1999 congress on Evolutionary Computation, Proceedings 1999.
[8] M. Dorigo, M. Birattari, and Thomas Stiitzle "Ant Colony Optimization- Artificial Ants as a Computational Intelligence Technique" IRIDIA Technical Report Series: TR/IRIDIA/2006-023.
[9] Jing Yang, Mai Xu, Wei Zhao and Baogue Xu, " A Multipath Routing Protocol Based on Clustering and Ant Colony Optimization for Wireless Sensor Networks", Sensors ISSN 1424-8220, 10,4521-4540, doi : 10.3390/s 100504521' 2010.
[10] S. Tanachaiwiwat, P. Dave, R. Bhindwale and A. Heimy, "Location-centric Isolation of Misbehavior and Trust Routing in Energy Constrained Sensor Netwoks", IEEE International Conference on Performance Computing and Communications, 2004.
[11] Mani Zarei, Amir Msoud Rahmani, Avesta Sasan, Mohammad Teshnehlab, "Fuzzy based trust estimation for congestion control in wireless sensor networks", 2009 International Conference on Intelligent Networking and Collaborative Systems.
[12] Mani Zarei, Amir Msoud Rahmani, Razieh Farazkish, Sara Zahirnia,"FCCTF: Fairness Congestion Control for a distrustful wireless sensor network using Fuzzy logic", 2010 10th International Conference on Hybrid Intelligent Systems.
[13] A.Chakraborty, S.Ganguly, M.K.Naskar, A.Karmakar, "A Trust Based Fuzzy Algorithm for Congestion Control in Wireless Multimedia Sensor Networks (TFCC)", proceeding of 2nd International Conference, ICIEV , Dhaka, Bangladesh, 2013.
[14] W. R .Heinzelman, A. Chandrakasan and H. Balakrishnan, "Energy Efficient Communication Protocol for Wireless Microsensor Networks" , Proceedings of the 33 Hawaii International Conference on System Sciences 2000.